\documentclass[preprint,aps,prl,showpacs]{revtex4}
\usepackage{graphicx}
\usepackage{dcolumn}
\usepackage{bm}
\usepackage{color}

\begin{document}

\preprint{APS/123-QED}

\title{Anomalous Lattice Parameter of Magnetic Semiconductor Alloys}

\author{Cl\'{o}vis Caetano$^1$}
\author{Luiz G. Ferreira$^2$}
\author{Marcelo Marques$^1$}
\author{Lara K. Teles$^1$}\email{lkteles@ita.br}

\affiliation{$^1$ Instituto
Tecnol\'{o}gico de Aeron\'{a}utica, 12228-900 S\~{a}o Jos\'{e} dos
Campos, SP, Brazil\\
$^2$ Instituto de F\'{i}sica, Universidade de S\~{a}o
Paulo, CP 66318, 05315-970 S\~{a}o Paulo, SP, Brazil
}%

\date{\today}

\begin{abstract}
The addition of transition metals (TM) to III-V semiconductors radically changes their electronic, magnetic and structural properties. In contrast to the conventional semiconductor alloys, the lattice parameter in magnetic semiconductor alloys, including the ones with diluted concentration (the diluted magnetic semiconductors - DMS), cannot be determined uniquely from the composition.
By using first-principles calculations, we find a direct correlation between the magnetic moment and the anion-TM bond lengths. We derive a simple formula that determines the lattice parameter of a particular magnetic semiconductor by considering both the composition and magnetic moment. The formula makes accurate predictions of the lattice parameter behavior of AlMnN, AlCrN, GaMnN, GaCrN, GaCrAs and GaMnAs alloys. This new dependence can explain some of the hitherto puzzling experimentally observed anomalies, as well as, stimulate other kind of theoretical and experimental investigations.

\end{abstract}

\pacs{75.50.Pp, 61.66.-f, 61.66.Dk, 71.20.Nr, 71.20.Be}

\maketitle

In the last years, the field of spin electronics or spintronics, which aims the use of the spin carriers in addition to the charge for electronic devices, has emerged as the new frontier for the future integrated circuit devices technology. The suggested applications include, e.g., spin transistors, spin valves, spin-polarized light emitting
diodes, as well as, potential applications in the field of quantum cryptography and quantum computing \cite{Ronaldo,Zutic,Pearton}.
To implement this new technology, efficient techniques for spin injection,
manipulation and detection are necessary, making it crucial the development
of materials with the needed properties \cite{Zutic,Pearton}. Materials under intense investigation are the diluted magnetic semiconductors (DMS), in which a conventional semiconductor is doped with TM atoms \cite{Ohno}. The great advantage of DMS is to allow an interplay between magnetic and electronic properties, being structurally compatible with most semiconductors epitaxially grown. Primarily, great attention was paid to the arsenide-based DMS.
Nowadays, a large class of DMS has already been proposed, being extensively studied, both experimental
and theoretically \cite{Jungwirth}.

Among all properties of DMS, the structural are of fundamental importance. The greater part of the III-V semiconductors crystallizes in the cubic zincblende (zb) structure, only the nitrides, that under ambient conditions crystallize in the hexagonal wurtzite
(wz) structure. However, successful growth of zb-AlN, GaN, and InN
has been
reported \cite{Pearton}.  On the
other hand, the structure of nitride based TM is the tetragonally distorted NaCl structure, NiAs-like for MnAs, and MnP-like for CrAs.
For III-V films containing no more than few percent of TM, one expects
the structure to be that of the host. Moreover, there has been an effort to grow these binary compounds in the zb structure using molecular beam
epitaxy. Recently, it was reported the growth of zb-MnAs \cite{Jeon}, and zb-CrAs \cite{Akinaga}. The formation of these binary compounds with the desirable zb structure is difficult because of inherent growth problems \cite{Jeon,Katsnelson}. Thus, their lattice constant in zb structure are not well established yet, and further experimental investigations are still needed.

In a pioneer work,  Ma\v{s}ek \textit{et al.}, by using a density functional theory (DFT) calculation, related the lattice parameter behavior
with the quality of the samples, by taking into account
the incorporation of TM in the interstitial positions \cite{Masek}. Afterwards, the lattice parameter behavior (increasing with the TM concentration) was taken as a fingerprint of the presence of interstitial TM, as well as, arsenic antisites in the GaMnAs samples. Furthermore, once the lattice constant can be precisely measured by X-ray diffraction, if one knows its relation with the composition of the alloy, the latter is directly determined, and in DMS the accurate determination of the alloy composition is of crucial importance because the TM content strongly affects the Curie temperature \cite{Dietl}. However, it was pointed out that clearly exists a phenomenological correlation between $a$ and $x$, but it is quite complex and not really understood \cite{Kudelska}. Moreover, it was experimentally observed that the GaMnAs is an example of a system which does not obey Vegard's law \cite{Schott}.

 Deviations from Vegard's law are derived from many factors, as electronic and/or magnetic
effects, as already observed for metallic alloys \cite{Shiga,Niculescu,Bombardi}. The exchange integrals in the Heisenberg model,
describing the super exchange interactions between magnetic ions in
DMS are strongly dependent on the lattice parameter \cite{Szuszkiwicz}. An important
property  for a material to be useful for
 spintronics  is its possible ferromagnetism.  Therefore,
 it should be of great interest, for a better understanding of the
magnetic properties of DMS, a rigorous study of the effect of TM impurities in the
structural properties of the magnetic semiconductors. Moreover, it should be very useful to have a quantitative relation between the lattice parameter and the magnetic properties, which, up to our knowledge, was not addressed until now for magnetic semiconductor alloys, nor for its dilute concentration.

In this work, we present a theoretical study of
structural, electronic, and magnetic properties of zb AlMnN, AlCrN, GaMnN, GaCrN, GaMnAs and GaCrAs alloys, considering the whole range of composition.
In particular, we focus our attention on the lattice parameter, analyze its relation to the other properties, and derive a relation between the lattice parameter and the average magnetic moment. Moreover, we also analyze how the behavior of lattice parameter is related to the electronic transition from the half-metal to the metal phase.

We used the frozen-core projector-augmented wave method as
implemented in the ``Vienna ab-initio simulation package'' (VASP-PAW
code) within a spin-polarized DFT \cite{VASP}.  We adopted the generalized gradient approximation for the
exchange-correlation potential in the version proposed by Wang and
Perdew \cite{Perdew}. It has been demonstrated that this method is able to successfully
describe the physical properties of several DMSs \cite{Marques,Jungwirth}. In order to calculate the ferromagnetic (FM) state, 16-atom supercells have been used, and varied the
number and the position of TM atoms (substitutional) in the cation sublattice, which results in 16 configurations not related by symmetry.  The lattice constant was optimized by total energy minimization for each configuration and the atomic positions were relaxed. In order to discuss the properties of the alloys we made averages over the configurations with the same composition $x$.

\begin{figure}
\includegraphics[width=\columnwidth]{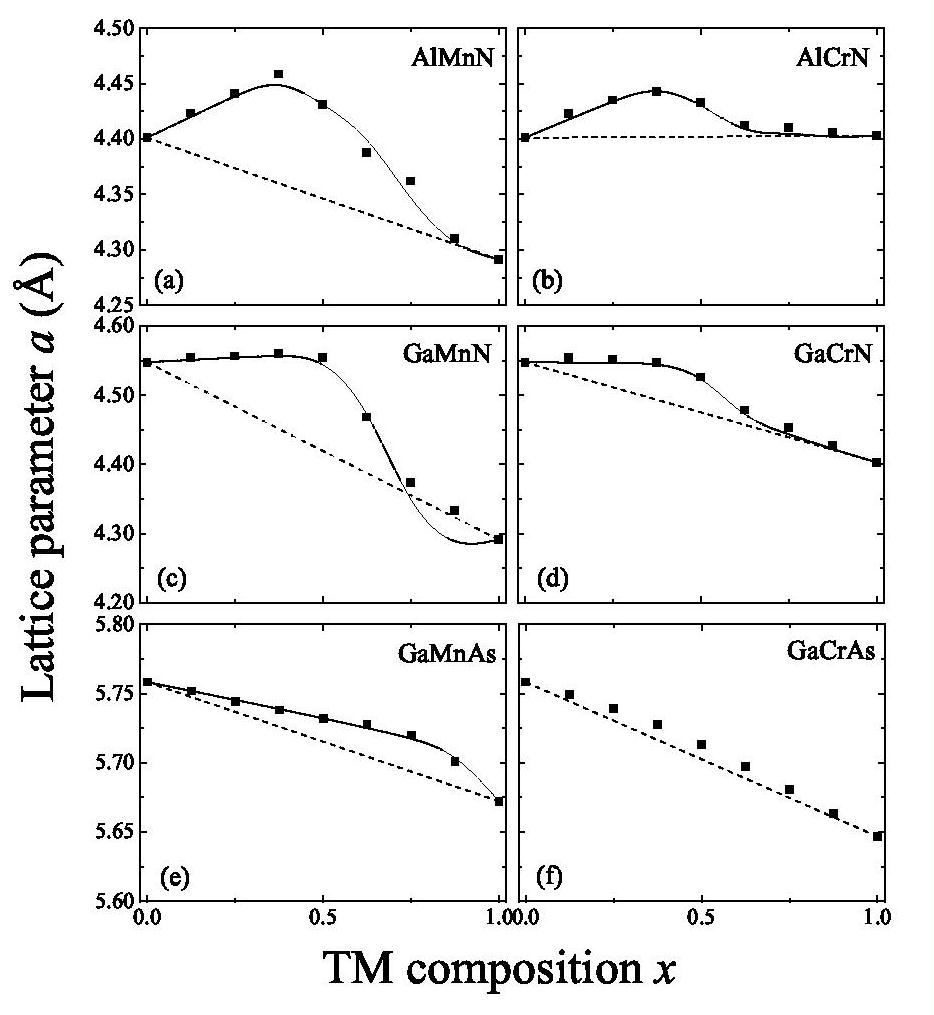}
\caption{\label{lattice_todos} Equilibrium lattice parameters (squares)
as function of the TM composition for the FM AlMnN, AlCrN, GaMnN, GaCrN, GaMnAs and GaCrAs. The solid line represents the fit according to Eq. \ref{a_x}, and the dashed line the lattice parameter behavior according to the Vegard's law.}
\end{figure}

Fig. \ref{lattice_todos} depicts the results for the lattice
parameter of AlMnN, AlCrN, GaMnN, GaCrN, GaMnAs and GaCrAs for the whole range of TM composition. By using the values of the optimized lattice
of FM MnN, CrN, and MnAs we note that Vegard's law predicts values in
complete disagreement with those observed. We observe an almost linear first increase in comparison to the Vegard's law
followed by a fast decrease of the lattice constant with the Cr and
Mn content. The case of GaCrAs is an exception that will be explained later.  This behavior is in agreement with other few available theoretical results \cite{Masek,Shi,Kanoun}, as well as, with experimental
data obtained by X-ray diffraction for AlMnN, AlCrN \cite{Sato2007,Zhang} and GaMnN \cite{Song,Leite} in the
wz structure. Here, it is worth
to point out that the results
for wz and zb should be very similar. We note that the increase of the lattice parameter is here due to the substitutional TM, not the interstitial as in Ref. \cite{Masek}. In the case of GaMnAs, it is experimentally observed an increase of the lattice constant with the Mn concentration for low Mn content \cite{Kudelska}, which is in disagreement with our prediction. In this case, { several other factors can be considered as the reason for this increasing, the interstitial atoms, As antisites, the presence of defects due to the growth at low temperatures, and deviations from stoichiometry \cite{Masek,Schott,Kudelska}. }

\begin{figure}
\includegraphics[width=\columnwidth]{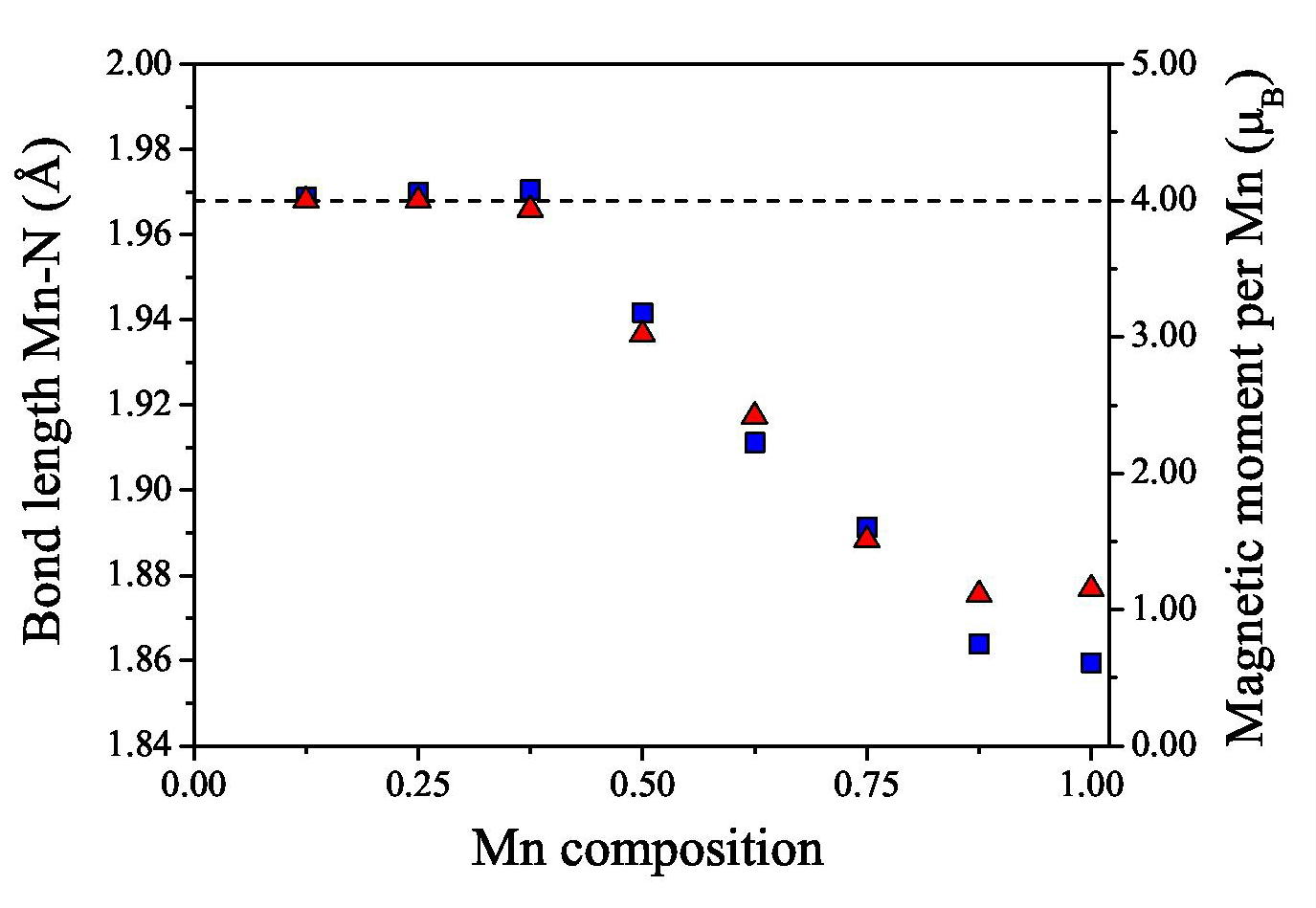}
\caption{\label{mag_bond_AlMnN}(color online) Total magnetic moment (red triangles)
and average bond-length (blue squares) for the ferromagnetic state of AlMnN as a function of Mn composition. The
dashed line represents the integer value 4.0$\mu_{B}$.}
\end{figure}

\begin{figure}
\includegraphics[width=\columnwidth]{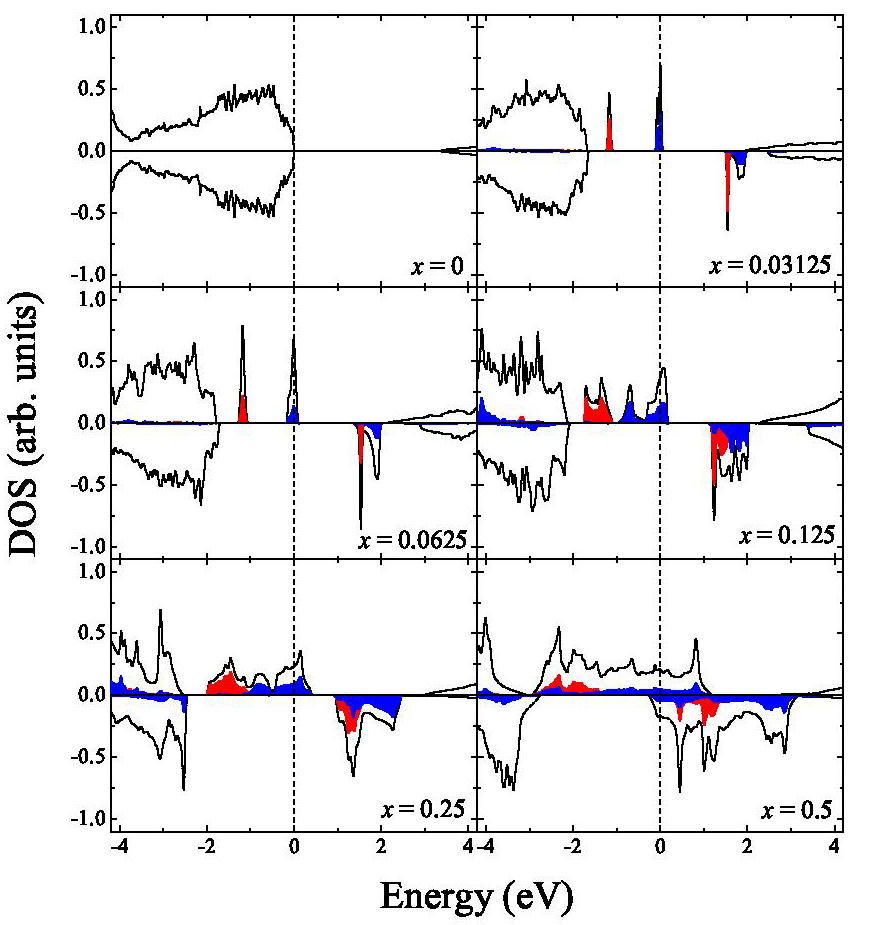}
\caption{\label{DOS} (color online) Total spin resolved density of
states of AlMnN for different Mn compositions (solid lines). The
partial DOS of $3d$ electrons, $e$ (red shaded) and $t_{2}$ (blue
shaded) states, of the Mn atoms is also shown. The vertical dashed
lines denote the positions of the Fermi level. }
\end{figure}

 As Vegard's law is a
consequence of the size difference between the host atoms and
the dopant, we infer that the observed deviation must result
from an electronic and/or magnetic effect. This kind of deviation has been
studied for solid solutions, but only for metal alloys \cite{Shiga,Niculescu,Bombardi}.
In order to investigate this, we analyze the nearest-neighbor distance behavior with the TM composition, which is depicted in Fig. \ref{mag_bond_AlMnN} for the AlMnN. Although the AlN lattice parameter suffers
an expansion when doped with a low concentration of TM, the bond
lengths TM-N remain nearly constant below a composition of
$\sim$30\%  TM. Another characteristic that we can note is the
correlation between the bond length and the total magnetic moment,
i.e., there exists an interdependence between the structural and
magnetic properties in those alloys. For low TM concentrations
the magnetic moment remains an integer, an evidence that the material
is half-metallic \cite{Katsnelson}, which is consistent with the null
density of states (DOS) for the minority spin-component, as shown for AlMnN in Fig.
\ref{DOS} (in order to analyze the DOS for low TM-content in more detail, we also used 64-atom supercells). When the TM composition increases, a transition
from the half-metallic to metallic behavior happens, as it can be
observed by the non-integer total magnetic moment, and by the DOS shown in Fig. \ref{DOS}

By taking into account this correlation between bond-length and the magnetic moment, we estimate the contribution from magnetization to the variation of the lattice parameter as follows.
 The Vegard's law is given by:
\begin{equation}
a_{\mbox{\scriptsize{(III,TM)V}}}(x)=(1-x)a_{\mbox{\scriptsize{III-V}}}+xa_{\mbox{\scriptsize{TM-V}}},
\end{equation}
where $a_{\mbox{\scriptsize{III-V}}}$ and $a_{\mbox{\scriptsize{TM-V}}}$ are the lattice parameters of the binaries, with III and V elements of column 3A and 5A from periodic table, respectively. For a zincblende structure, we can rewrite the equation above by expressing $a_{\mbox{\scriptsize{TM-V}}}$ in terms of the nearest-neighbor distance $d_{\mbox{\scriptsize{TM-V}}}$,
\begin{equation}\label{a_x_d}
a_{\mbox{\scriptsize{(III,TM)V}}}(x)=(1-x)a_{\mbox{\scriptsize{III-V}}}+x\frac{4}{\sqrt{3}}d_{\mbox{\scriptsize{TM-V}}}.
\end{equation}
For semiconductor alloys, the anion-cation bond-lengths do not converge to a single value, but instead remain close, throughout the composition range, to their respective values in the pure binary compounds \cite{Caetano}. This is not the case for magnetic semiconductors. Here, not only $d_{\mbox{\scriptsize{TM-V}}}$ dramatically  changes with the TM content, but also has a direct and approximately linear relation with the magnetic moment as
\begin{equation}
d_{\mbox{\scriptsize{TM-V}}}(x)=\alpha+\beta \mu(x),
\end{equation}
while the bond-length of the III-V remains approximately constant {($d_{Al-N} \simeq$ 1.90 \AA, $d_{Ga-N} \simeq$ 1.97 \AA, and $d_{Ga-As} \simeq$ 2.49 \AA)}. So that by generalizing Eq. \ref{a_x_d}
\begin{equation}
a_{\mbox{\scriptsize{(III,TM)V}}}(x)=(1-x)a_{\mbox{\scriptsize{III-V}}}+x\frac{4}{\sqrt{3}}(\alpha+\beta \mu(x)),
\end{equation}
from which follows immediately
\begin{equation}\label{a_x}
a_{\mbox{\scriptsize{(III,TM)V}}}(x)=(1-x)a_{\mbox{\scriptsize{III-V}}}+xa_{\mbox{\scriptsize{TM-V}}}+Cx(\mu(x)-\mu_{\mbox{\scriptsize{TM-V}}}),
\end{equation}
where the first part is exactly the Vegard's rule, and the last term represents the magnetic influence on lattice parameter, with $C = \frac{4\beta}{\sqrt{3}}$ a constant related to a magnetoelasticity between the TM and the anion, with unit of \AA/$\mu_{B}$, and $\mu_{\mbox{\scriptsize{TM-V}}}$ is the magnetic
moment per TM atom for the endpoint binary TM-V. The parameters $C$ thus obtained were all about 0.1 \AA/$\mu_{B}$. Explicitly they were 0.0914, 0.1199, 0.0992, 0.1307 and 0.1401 \AA/$\mu_{B}$ for AlMnN, AlCrN, GaMnN, GaCrN, and GaMnAs, respectively. A similar relation between the magnetic moment and the lattice parameter was already obtained a long time ago by Shiga for the 3d transition metal alloys Fe-Co and Fe-Ni \cite{Shiga}, and Niculescu \textit{et al.} for FeCoSi alloys \cite{Niculescu}.  The results for the equilibrium lattice parameter and the one obtained through the expression above are presented in Fig. \ref{lattice_todos}. We can observe that the suggested relation for the lattice parameter fits very well the calculated equilibrium lattice parameters. In the following, we give an interpretation for the lattice parameter behavior, and consequently the suggested equation.


By analyzing the electronic structure for the zb alloys, the TM \textit{d} ion levels are split by the tetrahedral crystal field into $t_2$ and $e$. Exchange interactions further
split these levels into spin-up and spin-down levels. The $t_2$ levels on the Mn atom may hybridize with the
levels with the same symmetry the anion \textit{p}-states, and can take part in the bond formation, while the $e$ levels have no other states available for significant
coupling, not taking part in bond formation.  The introduction of TM impurities on the host creates a spin-polarized impurity band, and the Fermi level lies on it, until the TM concentration achieves a value for which the alloy becomes a metal.

Let us first analyze the case of low TM content, as shown in Fig. \ref{DOS} for AlMnN. In this case the spin-polarized band is very narrow with almost no dispersion, meaning that { the overlap between adjacent Mn-induced levels is small, and the hybridization is not significant, since in this case the level is in the middle of the gap}. Thus the electrons on these levels are localized, which reduces the bond strength, resulting in an increase of the bond-length of the TM-anion. This is why the TM-anion bond-length is enlarged in comparison to the bond lengths of the metals MnN, MnAs and CrN, where all the outer electrons do participate in the bond formation, resulting in a stronger bond and reduced bond-length.   Therefore, for low TM content the alloy maintains its semiconductor character and the bond-lengths remain close to the enlarged TM-anion bond-lengths.  At the same time, as the bond-length remains constant, so does the magnetic moment, and the alloy is half-metal. Now, as the TM content increases there is a transition from half-metal to metal phase, and there is not an impurity level anymore, nor a semiconductor character for the bond-lengths, but a band with $d$ character, where the TM orbitals are overlapping. This overlapping decreases the kinetic energy, since the $d$-electrons become more itinerant and less localized. With the delocalization and decreasing of the kinetic energy, the bonding energy increases and the lattice parameter decreases. At the same time occurs the transition from localized to delocalized state, { the minority spin states becomes populated}, leading to a reduction of the magnetic moment and of the lattice parameter.

Finally, in order to give support to our conclusions, it should be useful to obtain the lattice parameter relation for an alloy which is half metal in its whole range of composition. This is the case of the GaCrAs alloy. In this case according to our proposition the lattice parameter should follow the Vegard's rule, since the magnetic moment of GaCrAs is the same in the whole range of composition, and equal to 3.0 $\mu_B$. Fig. \ref{lattice_todos}(f) depicts the result for the lattice constant behavior of this alloy. We can observe that there is no transition of behavior as for the other alloys, and the lattice parameter nearly follows the Vegard's law with a small bowing parameter of 0.04 \AA. Therefore corroborating our predictions.

In summary, in this work we have analyzed the lattice parameter
variation of cubic magnetic semiconductors of family III-V by using first-principles
calculations. We conclude that incorporation of TM impurities can significantly
affect the lattice parameters. A large deviation from Vegard's law was observed.  We noticed a correlation between the magnetic moment and the interatomic distances between the TM and its first-neighbor, which
lead us to conclude that there is an influence from the electronic/magnetic properties. Hopefully these studies on the correlation of the electronic and magnetic effects with the structural ones, will play an important role in the
understanding of the remaining questions and stimulate other kind of theoretical and experimental investigations, allowing a more thorough understanding of these enigmatic materials.
\par This work was supported by the Brazilian funding agencies CNPq and FAPESP (procs. 05/52231-0 and 06/05858-0).


\newpage

\end{document}